\documentclass[10pt,conference,twoside]{IEEEtran}
\usepackage{stfloats}
\usepackage{amsmath}
\usepackage{graphicx} %use graph format
\usepackage{amssymb}
\usepackage{amsfonts}
\usepackage{amsthm}
\usepackage{cite}
\usepackage{bm,comment}
\usepackage{algorithm}
\usepackage{algpseudocode}

\usepackage{subeqnarray}
\usepackage{subfigure}
\usepackage{multirow}
\usepackage{diagbox}
\usepackage{slashbox}
\usepackage{stfloats}
\usepackage{float}
\usepackage{color} 
\usepackage{cases}

\newtheorem{proposition}{Proposition}

\floatname{algorithm}{Algorithm}

\begin{document}
%\IEEEoverridecommandlockouts
\title{3D Beam Training in Terahertz Communication:\\ A Quadruple-UPA Architecture}
\author{\IEEEauthorblockN{Zhongbao Tian\IEEEauthorrefmark{1}, Zhi Chen\IEEEauthorrefmark{1}, and Boyu Ning\IEEEauthorrefmark{1}}\\
\IEEEauthorblockA{\IEEEauthorrefmark{1}National Key Laboratory of Science and Technology on Communications\\
University of Electronic Science and Technology of China (UESTC), Chengdu 611731, China\\
Emails: boydning@outlook.com;}
\thanks{This work was supported by the National Key R$\&$D Program of China under Grant 2018YFB1801500.}}
\maketitle

\begin{abstract}
In this paper, we propose a quadruple-UPA architecture for realizing the beamforming with wide coverage in terahertz (THz) communication. Considering the severe path loss suffered in the THz wave propagation, the pilot signals of conventional channel estimation strategies may not be effectively detected. In sight of this, we propose a fast 3D beam training strategy to jointly realize the angle estimation and the beamforming. Specifically, we first develop a novel hierarchical codebook that predefines some codewords (including narrow beams and wide beams) stage by stage. The narrow beams are designed as the beamforming solution candidates, whereas the wide beams are designed for reducing the training complexity. Then, we propose a 3D beam training procedure to find the optimal beamforming solution, i.e., the optimal narrow-beam pair, in a fast manner. Numerical results are presented to validate the effectiveness of the proposed scheme.

\end{abstract}
\IEEEpeerreviewmaketitle

\section{Introduction}
Terahertz (THz) communication is envisioned as a key wireless technology to alleviate the spectrum bottleneck and support high data rates in the future\cite{t1}. The THz band, ranging from $0.1$ THz to $10$ THz, supports huge transmission bandwidths with multiple transmission windows separated by the attenuation absorption peaks\cite{t2}. Despite some transmission windows are with bandwidth more than hundreds of GHz, they come at the cost of a very high propagation loss suffered from both spread path loss and molecular absorption\cite{t3}. To compensate the severe propagation loss, various technologies, e.g., massive or ultra-massive MIMO \cite{m1,m2}, coordinated multi-point transmission \cite{cm}, and intelligent reflecting surface\cite{b1,b2}, can be integrated in THz communications to provide effective multiplexing gains. In the THz massive MIMO systems, the transceiver and receiver equipped with large-scale antenna arrays can realize the directional communication with sufficient beam gain by dynamically controlling the amplitude and phase shifts on each element\cite{mm}. However, the coverage of the transmit/receive signal on one array is limited, which significantly reduces the flexibility of the beam control\cite{ca}. For example, the uniform planar array (UPA) covers only $\pm \frac{\pi }{2}$ both azimuth and elevation to its boresight, i.e., in the front range of the face of the array. 

In this paper, we propose a quadruple-UPA architecture to realize the 3D beamforming with wide coverage in THz communication. To be exact, it covers omni-direction in azimuth and $\frac{\pi }{2}$ range in elevation, and each UPA only covers $\frac{\pi }{2}$ range in both azimuth and elevation. Beyond the wide coverage, the quadruple-UPA architecture also has the following advantages. 1) Considering the high directivity of THz waves, by expanding the beam space to 3D, this architecture has greater potential for high-quality THz communications, e.g., effective coexistence between drones and cellular networks \cite{mei}. 2) Each UPA in this architecture can achieve a higher array gain than conventional one, by using the directional antenna element tailored for its coverage. Regarding the quadruple-UPA architecture, we further propose a low-complexity 3D beam training strategy to jointly realize the angle estimation and the beamforming. Specifically, we first propose a novel hierarchical codebook that predefines some codewords (including narrow beams and wide beams) stage by stage. The narrow beams act as the beamforming solution candidates, which determines the overall training performance.  The wide beams divide the coverage range into separate parts, which are used for reducing the training complexity. Compared to the existing training codebooks \cite{wide1,wide3,wide4,ywang}, our proposed narrow beams guarantee the maximum worst-case performance, and our proposed wide beams have a less dead zone in the beam training. With the proposed hierarchical codebook, we further develop a 3D beam training procedure to find the optimal beamforming solution, i.e., the optimal narrow-beam pair. This procedure incurs much lower training complexity compared to exhaustively searching the narrow beams.

\begin{figure}[t]
\centering
\includegraphics[width=2.6in]{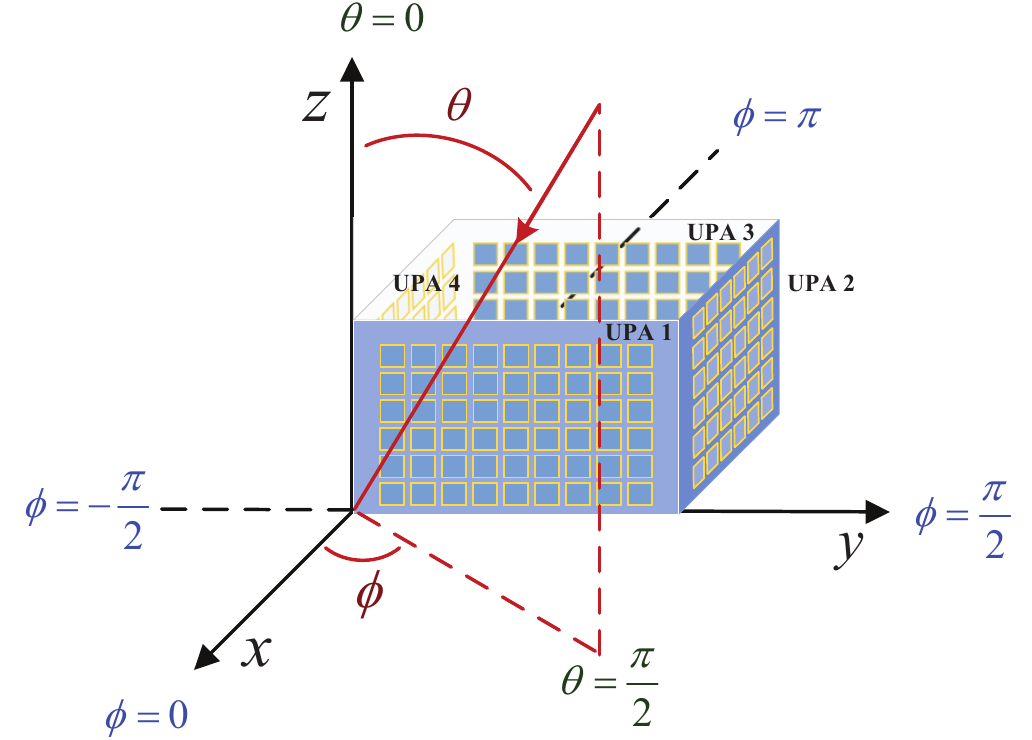}
\caption{Four identical UPAs equipped at the transmitter and the receiver.}\label{mo}
\vspace{-12pt}
\end{figure}

\section{System and Problem Descriptions}
\subsection{System Model}
\begin{figure}[t]
\centering
\includegraphics[width=2.7in]{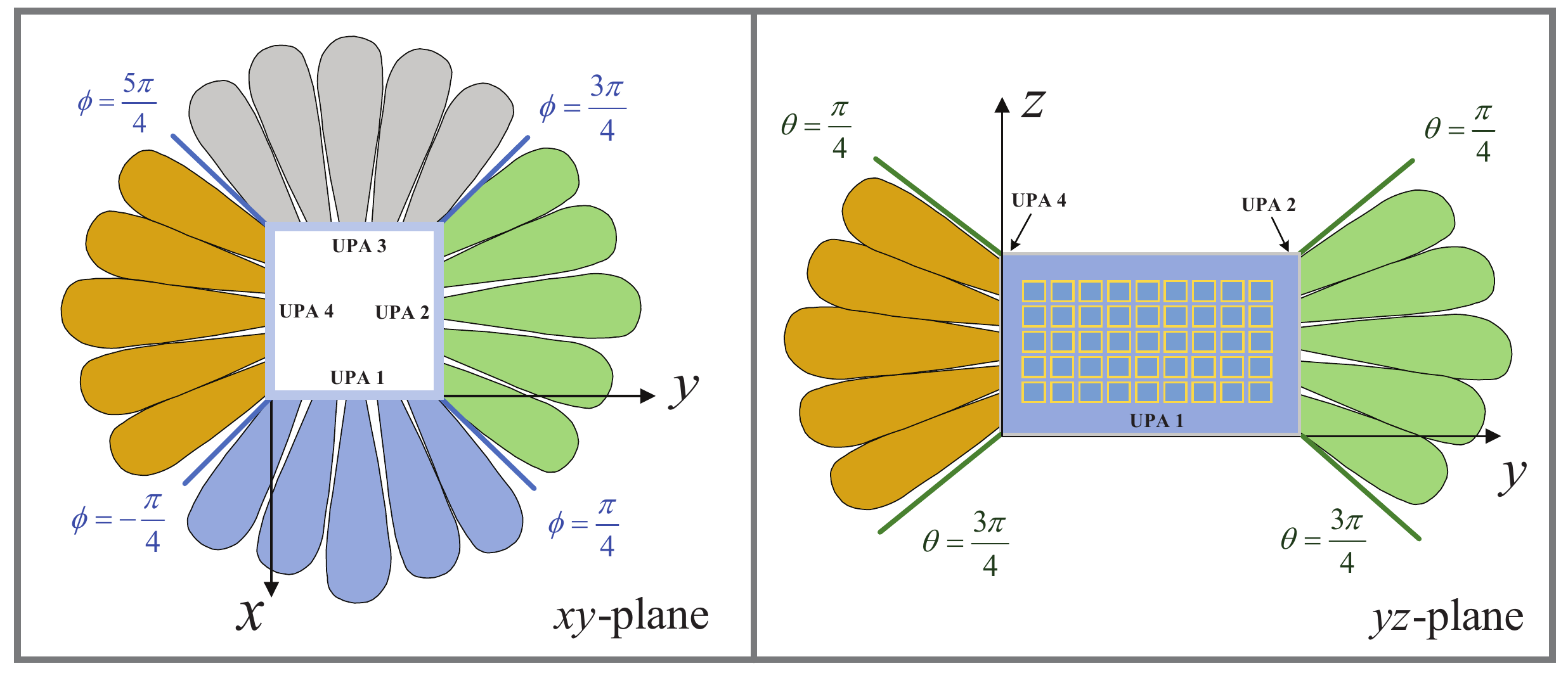}
\caption{Transmit and receive ranges of the four UPAs on the $xy$-plane and $yz$-plane.}\label{mo2}
\vspace{-12pt}
\end{figure}

We consider a point-to-point THz communication system with four half-wave spaced UPAs, a quadruple-UPA architecture, equipped at the transmitter and receiver, respectively. Without loss of generality, we assume that both the transmitter and the receiver have the same architecture where four identical UPAs with $N_a$ elements are equipped around a cube. As shown in Fig. \ref{mo}, we use $x$, $y$, and $z$-axes to refer to the axes of the standard Cartesian coordinate system for all the UPAs. In the case of the $1$st UPA, with $N_y$ and $N_z$ elements on the $y$ and $z$-axes respectively ($N_a=N_y N_z$), the array response vector can be expressed in a conventional form\footnote{We assume the signal phase at the center of the UPA is zero.}, i.e.,
\begin{equation}
\begin{split}
&{{\bf{a}}_1}(\phi ,\theta ) \!=\! \frac{1}{{\sqrt {{N_a}} }}[{e^{j\pi [ - \frac{{({N_y} - 1)}}{2}\sin \phi \sin \theta  - \frac{{({N_z} - 1)}}{2}\cos \theta ]}},...,\\
&\;\;\;\;\;\;\;\;\;\;\;\;\;\;\;\;\quad\qquad{e^{j\pi ({n_y}\sin \phi \sin \theta  + {n_z}\cos \theta )}},...,\\
&\;\;\;\;\;\;\;\;\;\;\;\;\;\;\;\;\qquad\qquad{e^{j\pi [\frac{{({N_y} - 1)}}{2}\sin \phi \sin \theta  + \frac{{({N_z} - 1)}}{2}\cos \theta ]}}{]^T},
\end{split}
\end{equation} 
where $\phi$ and $\theta$ are the azimuth angle to $x$-axis and the elevation angle to $z$-axis respectively, ${n_y} =  - \frac{{({N_y} - 1)}}{2} + 1, - \frac{{({N_y} - 1)}}{2} + 2,...,\frac{{({N_y} - 1)}}{2} - 1$, ${n_z} =  - \frac{{({N_z} - 1)}}{2} + 1, - \frac{{({N_z} - 1)}}{2} + 2,...,\frac{{({N_z} - 1)}}{2} - 1$. Note that the perpendicular direction of the $k$th array is $(\frac{{(k - 1)\pi }}{2},\frac{\pi }{2})$, thereby the response vector of the $k$th array can be written as 
\begin{equation}\label{array}
\begin{split}
&{{\bf{a}}_k}(\phi ,\theta ) \!=\! \frac{1}{{\sqrt {{N_a}} }}[{e^{j\pi \left[ { - \frac{{({N_y} - 1)}}{2}\sin \left( {\phi  - \frac{{(k - 1)\pi }}{2}} \right)\sin \theta  +  - \frac{{({N_z} - 1)}}{2}\cos \theta } \right]}},\\
&\;\;\;\;\;\;\;\;\qquad...,{e^{j\pi \left[ {{n_y}\sin \left( {\phi  - \frac{{(k - 1)\pi }}{2}} \right)\sin \theta  + {n_z}\cos \theta } \right]}},\\
&\;\;\;\;\;\;\;\;\quad\qquad...,{e^{j\pi \left[ {\frac{{({N_y} - 1)}}{2}\sin \left( {\phi  - \frac{{(k - 1)\pi }}{2}} \right)\sin \theta  + \frac{{({N_z} - 1)}}{2}\cos \theta } \right]}}{]^T}.
\end{split}
\end{equation}
To provide omni-directional communication with adequate array gains, four UPAs are tailored for beamforming in four different space ranges by using the directional antenna elements respectively. As shown in Fig. \ref{mo2}, each array is dedicated to transmitting and receiving signals \emph{only} in the range within $ \pm \frac{\pi }{4}$ to the perpendicular direction of the array, in both azimuth and elevation space. As such, the transmit/receive range of $k$th array is denoted by
\begin{equation}\label{omi}
{\Omega _k}=\left\{ {({\phi _k},{\theta _k})\left| {\begin{array}{*{20}{c}}
{ - \frac{\pi }{4} + (k - 1)\frac{\pi }{2} \le {\phi _k} \le \frac{\pi }{4} + (k - 1)\frac{\pi }{2},}\\
{\frac{\pi }{4} \le {\theta _k} \le \frac{{3\pi }}{4}.}
\end{array}} \right.} \right\}.
\end{equation}
Let $s$ denote a transmitted symbol with unit power to the $k$th transmit UPA, the processed received signal from the $m$th receive UPA can be expressed as
\begin{equation}\label{rec}
{y_{k,m}} = \sqrt P {\bf{w}}_m^H {{{\bf{H}}_{k,m}}{{\bf{f}}_k}s}  + {\bf{w}}_m^H{\bf{n}},
\end{equation}
where $P$ represents the transmit power, ${{\bf{H}}_{k,m}}\in {\mathbb{C}^{N_a \times N_a}}$ is the channel matrix between the $k$th transmit UPA and $m$th receive UPA, $s$ is the data symbol,  ${\bf{f}}_k \in {\mathbb{C}^{N_a \times 1}}$ (resp. ${\bf{w}}_m \in {\mathbb{C}^{N_a \times 1}}$) is the normalized beamforming precoder (resp. decoder) at $k$th (resp. $m$th) UPA, and ${\bf{n}} \sim \mathcal{CN}({\bf{0}},\sigma^2{\bf{I}})$ is the zero-mean additive Gaussian noise with power $\sigma^2$.

\subsection{Problem Statement}
To enable a reliable THz communication, the precoder ${\bf{f}}_k$
at the $k$th transmit UPA and the decoder ${\bf{w}}_m$ at the $m$th receive UPA are needed to be optimized to maximize the decoding SNR, which is equivalent to solving the following problem
\begin{equation}
\begin{split}
\{ {\bf{w}}_m^{{\rm{opt}}},{\bf{f}}_k^{{\rm{opt}}}\}  = \arg &\max {\left| {{\bf{w}}_m^H{{\bf{H}}_{k,m}}{{\bf{f}}_k}} \right|^2}\\
&{\rm{s}}{\rm{.t}}{\rm{.}}\;\;{\left\| {{{\bf{f}}_k}} \right\|^2} \le 1,\;\;{\left\| {{{\bf{w}}_m}} \right\|^2} \le 1.
\end{split}
\end{equation}
Provided that ${{\bf{H}}_{k,m}}$ is perfectly known at the transmitter and receiver, the optimal precoder ${\bf{f}}_k^{{\rm{opt}}}$ and the decoder ${\bf{w}}_m^{{\rm{opt}}}$ can be easily derived by applying the singular value decomposition (SVD) on ${{\bf{H}}_{k,m}}$. However, the conventional channel estimation methods tailored for micro-wave and millimeter-wave communications may not apply to THz system since their pilot signals suffer from severe path loss in THz channels and can not be effectively detected by the receiver before efficient beamforming. Considering this fact, we need to find ${\bf{f}}_k^{{\rm{opt}}}$ and ${\bf{w}}_m^{{\rm{opt}}}$ by testing the precoder-decoder pairs predefined in a codebook, without the channel state information. From this perspective, THz channel estimation relies on effective beamforming. Note that the severe path loss also significantly limits scattering in THz communication, where the gain of the NLoS paths is much lower (more than 20dB) than that of the LoS counterpart. Therefore, in this paper, we mainly consider the LoS component in the beam training, i.e., 
\begin{equation}\label{tran}
\begin{split}
\{ {\bf{w}}_m^{{\rm{opt}}},{\bf{f}}_k^{{\rm{opt}}}\}  = &\arg \max {\left| {{\bf{w}}_m^H{\bf{H}}_{k,m}^{{\rm{LoS}}}{{\bf{f}}_k}} \right|^2}\\
&{\rm{s}}.{\rm{t}}.\;\;{\bf{H}}_{k,m}^{{\rm{LoS}}} = {G_t}{G_r}{\alpha _{\rm{L}}}{{\bf{a}}_m}({\phi _r},{\theta _r}){{\bf{a}}_k}{({\phi _t},{\theta _t})^H},\\
&\;\;\;\;\;\quad{{\bf{f}}_k} \in {{\cal F}_k},{{\bf{w}}_m} \in {{\cal W}_m},
\end{split}
\end{equation}
where ${\alpha}_{{\rm{L}}}$ is the complex gain of the LoS path, ${{\bf{a}}_k}({\phi _t},{\theta _t})$ and ${{\bf{a}}_m}({\phi _r},{\theta _r})$ are the normalized transmit and receive array response vectors which both follow the definition given in (\ref{array}). $ ({\phi _t},{\theta _t})$ and $({\phi _r},{\theta _r})$  are the LoS path's azimuth and elevation angles of departure and arrival (AoDs/AoAs), respectively. $\mathcal{F}_k$ and $\mathcal{W}_m$ are predefined codebooks for $k$th precoder and $m$th decoder, respectively. In fact, if the AoD and AoA of the LoS path are within the ranges of the $i$th UPA at the transmitter and the $j$th UPA at the receiver, other UPA pairs can not establish an effective communication link. This allows us to find a valid UPA pairs first, and then only considering the beam training problem on this pair.

\section{Beam Training Strategy}
In this section, we first introduce an exhaustive beam training method that can be applied in our considered system. Then, we further develop a more efficient approach to achieve a better performance-complexity trade-off. 
\subsection{Exhaustive Beam Training}
Based on (\ref{tran}), the beam training problem is equivalent to
\begin{equation}\label{ori}
\begin{split}
\{ {\bf{w}}_m^{{\rm{opt}}},{\bf{f}}_k^{{\rm{opt}}}\}  = \arg &\max {\left| {{\bf{w}}_m^H\underbrace {{{\bf{a}}_m}({\phi _r},{\theta _r}){{\bf{a}}_k}{{({\phi _t},{\theta _t})}^H}}_{{\rm{can}}\;{\rm{not}}\;{\rm{be}}\;{\rm{obtained}}}{{\bf{f}}_k}} \right|^2}\\
&{\rm{s}}{\rm{.t}}{\rm{.}}\;{{\bf{f}}_k} \in \mathcal{F}_k,{{\bf{w}}_m} \in \mathcal{W}_m,
\end{split}
\end{equation}
One can notice that without the codebook constraint, an optimal solution to (\ref{ori}) is given by $\{ {\bf{w}}_m^{{\rm{opt}}} = {{\bf{a}}_m}({\phi _r},{\theta _r}),{\bf{f}}_k^{{\rm{opt}}} = {{\bf{a}}_k}({\phi _t},{\theta _t})\} $. Since the optimal precoder and decoder follow the form of array response vector, a straightforward method to reach a desired solution is to traverse all precoder-decoder pairs from codebooks composed of array response vectors with different angles. This method is also referred to as \emph{exhaustive 3D beam training}\cite{ywang}. As for our considered system with four UPAs, the transmitter and receiver have $4N^2$ narrow beams on each. Thus, there are $16N^4$ beam pairs to be tested in the exhaustive beam training, which is quite time consuming when $N$ is large. 
\subsection{Hierarchical Codebook Design}\label{HCD}
Next, we propose a novel 3D hierarchical codebook to realize the beam training in a fast way, where the codewords in different stages represent beams with different beam width. Due to the common structure, we define $\mathcal{C}_k$ as the codebook for $k$th UPA of both the transmitter and the receiver.  Firstly, we design ${N^2} = {2^S}$ narrow beams which can cover $\Omega_k$ in union, where $S$ is the number of stages of our proposed hierarchical codebook. These narrow beams lie in the bottom stage, i.e., stage $S$, of the proposed hierarchical codebook and one narrow beam among will be selected as the optimal precoder/decoder after the hierarchical beam training. Thus, based on (\ref{ori}), all the narrow beams ought to follow the form of array response vector. As such, the design of the ${N^2}$ narrow beams is reduced to determining their directions, i.e., $\phi$ and $\theta$ in ${{\bf{a}}_k}(\phi ,\theta )$.  Assume that an optimal decoder is used in the beam training, based on (\ref{ori}), the received normalized decoding power is given by 
\begin{equation}
\begin{split}
&\mathop {\max }\limits_{{{\bf{f}}_k}} {\left| {{\bf{a}}_k}({\phi _t},{\theta _t})^H{{\bf{f}}_k} \right|^2}\\
&{\rm{s}}.{\rm{t}}.\;{{\bf{f}}_k} \in \mathcal{C}_k^S,
\end{split}
\end{equation}
where $\mathcal{C}_k^S$ represents the $N^2$ narrow beams to be designed in the stage $S$ of $\mathcal{C}_k$. Due to the randomness of the wireless channel (random ${\phi _t}$ and ${\theta _t}$), the quality of the codewords $\mathcal{C}_k^S$  can be judged by its one-side worst-case performance, i.e.,
 \begin{equation}\label{worst}
\begin{split}
{\eta _{{\rm{worst}}}}  = &\mathop {\min }\limits_{{\phi _t},{\theta _t}} \mathop {\max }\limits_{{{\bf{f}}_k}} {\left|  {{\bf{a}}_k}({\phi _t},{\theta _t})^H{{\bf{f}}_k} \right|^2}\\
&{\rm{s}}.{\rm{t}}.\;{{\bf{f}}_k} \in \mathcal{C}_k^S.
\end{split}
\end{equation}
Thus, we endeavour to design the directions for narrow beams of $\mathcal{C}_k^S$, to guarantee the highest worst-case performance. As a result, in our proposed hierarchical codebook, the $N^2$ narrow beams of $\mathcal{C}_k^S$ are given by 
\begin{subequations}\label{15o}
\begin{align}
C_k^S &= \left\{ {{{\bf{a}}_k}\left( {{\phi _n},{\theta _p}} \right)|n,p = 1,2,...,N} \right\},\;\\
{\phi _n} &= \arcsin \left( {\frac{{\sqrt 2 (2n - 1 - N)}}{{2N}}} \right) + \frac{{(k - 1)\pi }}{2},\label{15b}\\
{\theta _p} &= \arccos \left( {\frac{{\sqrt 2 (2p - 1 - N)}}{{2N}}} \right). \label{15c}
\end{align}
\end{subequations}
%It is worth noting that in $\mathcal{C}_k^S$, the azimuth angle $\phi_n$ is affiliated to the elevation angle $\theta_p$. Hence, we first design $N$ elevation angles via (\ref{15b}). For each elevation angle, we further design $N$ azimuth angles tailored for it via (\ref{15c}). Finally, we obtain the $N^2$ narrow beams for $\mathcal{C}_k^S$. 

\begin{proposition}  
If $N^2$ narrow beams (with $N$ azimuth angles times $N$ elevation angles) are adopted to cover $\Omega_k$ in union, the codewords proposed in (\ref{15o}) guarantees the highest worst-case performance (defined in (\ref{worst})), which is given by (normalized by the best-case performance) 
\begin{equation}\label{p1}
{\eta _{{\rm{worst}}}} = \frac{{\sin \left[ {{{\left( {\sqrt 2 {N_z}\pi } \right)} \mathord{\left/
 {\vphantom {{\left( {\sqrt 2 {N_z}\pi } \right)} {4N}}} \right.
 \kern-\nulldelimiterspace} {4N}}} \right]\sin \left[ {{{\left( {\sqrt 2 \beta {N_y}\pi } \right)} \mathord{\left/
 {\vphantom {{\left( {\sqrt 2 \beta {N_y}\pi } \right)} {4N}}} \right.
 \kern-\nulldelimiterspace} {4N}}} \right]}}{N_yN_z{\sin \left[ {{{\left( {\sqrt 2 \pi } \right)} \mathord{\left/
 {\vphantom {{\left( {\sqrt 2 \pi } \right)} {4N}}} \right.
 \kern-\nulldelimiterspace} {4N}}} \right]\sin \left[ {{{\sqrt 2 \beta \pi } \mathord{\left/
 {\vphantom {{\sqrt 2 \beta \pi } {4N}}} \right.
 \kern-\nulldelimiterspace} {4N}}} \right]}},
\end{equation}
where 
\begin{equation}\label{p2}
\beta  = \left\{ {\begin{split}
&{1,\quad\qquad\qquad\qquad{\rm{when}}\;N\;{\rm{is}}\;{\rm{odd}}}\\
&{\sin \Big( {\arccos \frac{{\sqrt 2 }}{{2N}}} \Big),\;\;{\rm{when}}\;N\;{\rm{is}}\;{\rm{even}}}
\end{split}} \right..
\end{equation}
\end{proposition}
\begin{IEEEproof}
The proof is omitted due to the page limit.
\end{IEEEproof}

It is worth mentioning that the advantage of the proposed hierarchical codebook is to reduce the time complexity while its performance is the same as that of exhaustively testing the $N^2$ narrow beams of $\mathcal{C}_k^S$. As for our considered system, we have ${N^2} = {2^S}$ which indicates that $N$ is even. Thus, Proposition 1 provides the guarantee of the normalized one-side worst-case performance of the proposed hierarchical codebook, which is given by (\ref{p1}) with $\beta=\sin \Big( {\arccos \frac{{\sqrt 2 }}{{2N}}} \Big)$. 

Next, we propose an approach to design the wide beams of upper stages, i.e., stage $s=0,1,...,S-1$, of the hierarchical codebook $\mathcal{C}_k$. Each wide beam in the stage $s$ covers two beams in the stage $s+1$ while the beam in stage $0$ covers the whole range of $\Omega_k$. As such, we have $2^s$ beams in the stage $s$. The codewords for wide beams are no longer array response vectors and we use $\bm{\omega} _i^{k,s}$ to represent the $i$th codeword in the stage $s$ of the hierarchical codebook $\mathcal{C}_k$. 

\begin{figure}[t]
\centering
\includegraphics[width=3.2in]{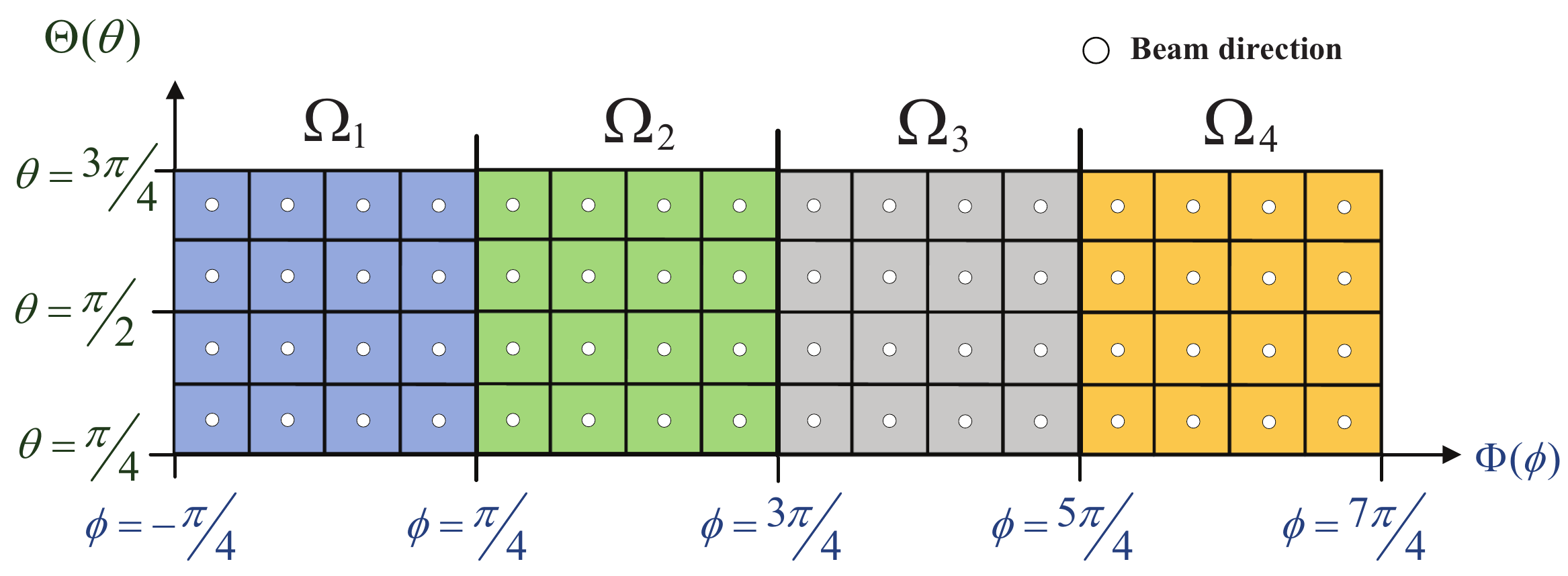}
\caption{An illustration of the range of the narrow beams shown on $\Theta (\theta )$ and $\Phi (\phi )$, where $N=4$.}\label{unir}
\vspace{-10pt}
\end{figure}

\begin{figure}[t]
\centering
\includegraphics[width=3in]{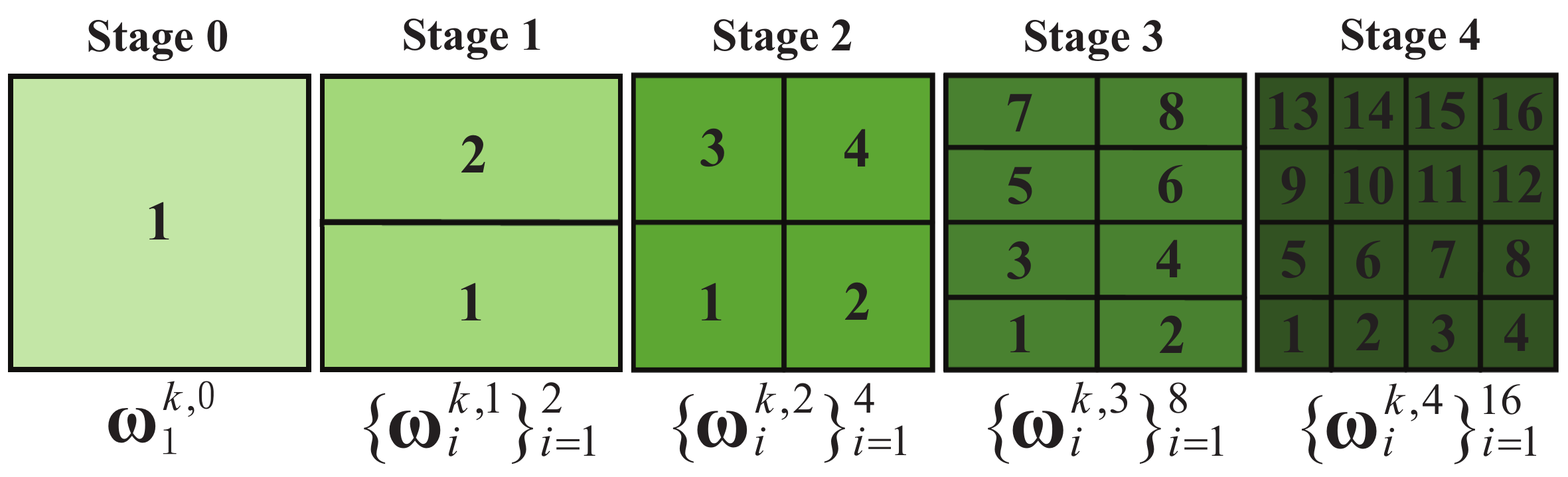}
\caption{Beams' distribution and their coverage in different stages, where $N^2=16$ and $S=4$.}\label{stg}
\vspace{-10pt}
\end{figure}

\begin{figure}[t]
\centering
\includegraphics[width=3in]{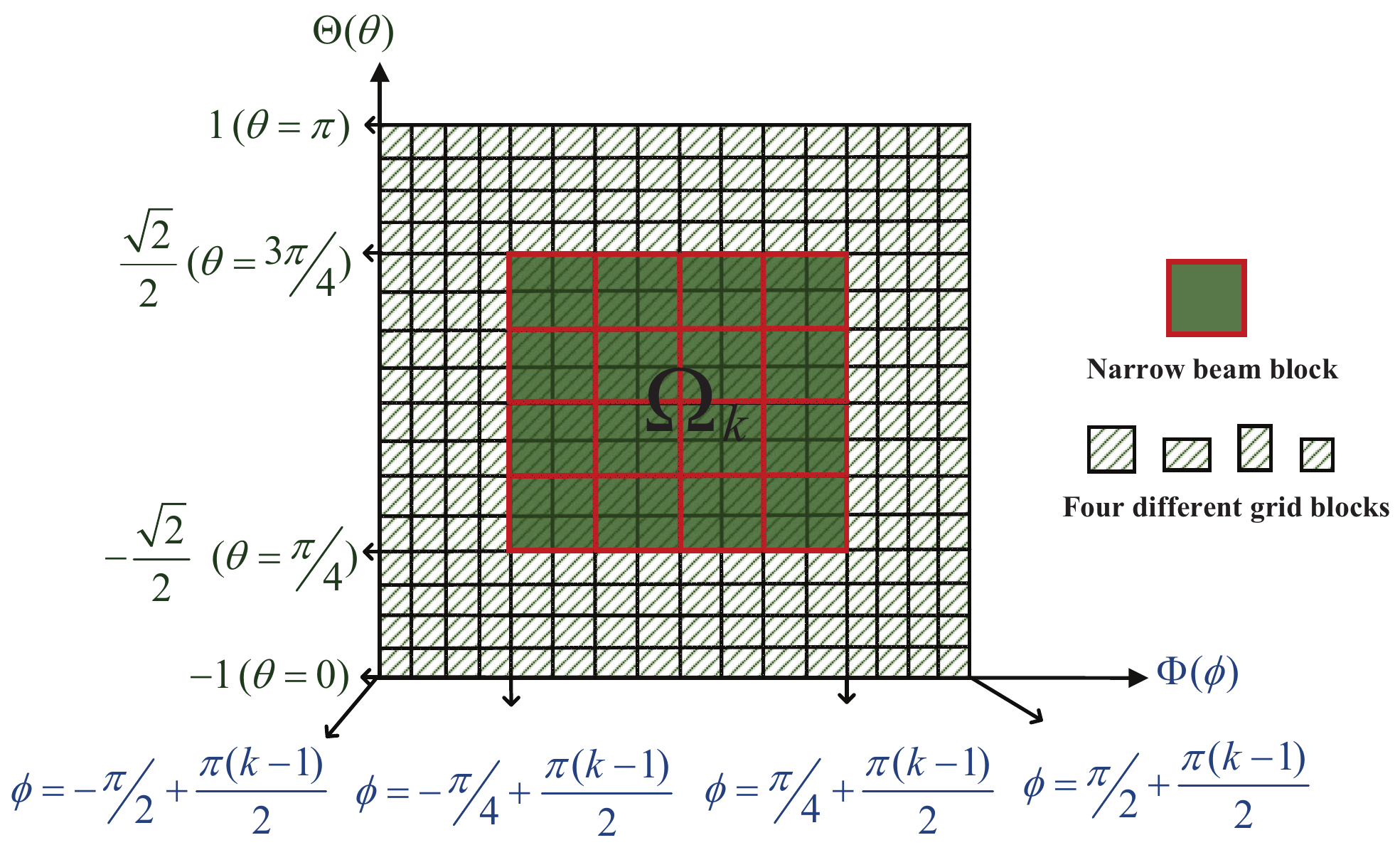}
\caption{A grid that represents all directions in front of the $k$th UPA, where $N=4$.}\label{gri}
\vspace{-10pt}
\end{figure}

As mentioned above, to design the wide beams in stage $S-1$, we need to know the coverage of the beams in stage $S$, i.e., the narrow beams. Note that the directions of the narrow beams are non-uniformly distributed in the 3D space, i.e., these beams have smaller coverage in the center of $\Omega_k$ but larger coverage around the edge of $\Omega_k$. For ease of illustrating their range, we define two functions as $\Theta (\theta ) =  - \cos \theta$ and $\Phi (\phi ) = \sin (\phi  - \frac{{\pi (k - 1)}}{2}) + \sqrt 2 (k - 1)$. By this means, as shown in Fig. {\ref{unir}}, the range of the narrow beams can be represented as squares on $\Theta (\theta )$ and $\Phi (\phi )$, where the beam direction is in the center of the square. Based on this representation, Fig. {\ref{stg}} shows our proposed beams' distribution as well as their coverage in different stages. 
According to the beams' distribution shown, the codewords of narrow beams are given by
\begin{equation}\label{narrow}
\begin{split}
&\bm{\omega} _i^{k,S} = {{\bf{a}}_k}({\phi _n},{\theta _p}),\;\;i = 1,2,...,{N^2},\\
&n = {\bmod _N}(i),\;\;p = {\rm{ceil}}(i/N),\;\;(\ref{15b}),\;\;(\ref{15c}).
\end{split}
\end{equation}
where $\bmod _N (i)$ returns the remainder after division of $i$ by $N$, and ${\rm{ceil}}(\cdot)$ returns the nearest integer greater than or equal to its argument. To develop the codewords of wide beams for $\mathcal{C}_k$, we have to construct a dense grid that represents all directions in front of the $k$th UPA, i.e.,  $\hat\Omega_k=\left\{ {(\phi ,\theta )|\phi  \in [ - \frac{\pi }{2} + \frac{{\pi (k - 1)}}{2},\frac{\pi }{2} + \frac{{\pi (k - 1)}}{2}],\theta  \in [0,\pi ]} \right\}$, which is larger than $\Omega_k$. As there are $N\times N$ narrow beams within range $\Omega_k$, as shown in Fig. {\ref{gri}}, we construct $2N\times 2N$ grid blocks within this range, and total $4N\times 4N$ grid blocks within $\hat\Omega_k$. We should mention that if the rest $4N\!\times\! 4N\!-\!2N\!\times\! 2N$ blocks each has the same size of that within $\Omega_k$, the total coverage is beyond $\hat\Omega_k$. Thus, we set them smaller and uniformly distributed on $\Theta (\theta )$ and $\Phi (\phi )$ to exactly cover  $\hat\Omega_k$. As such, the center directions of the grid blocks for $k$th UPA can be represented as 
\begin{small}
\begin{equation*}
\begin{split}
&(\phi _{{\rm{grid}}}^{k,j},\theta _{{\rm{grid}}}^{k,l})\;\;{\rm{with}}\;j= 1,2,...,4N,\;{\rm{and}}\;l= 1,2,...,4N,\\
&\phi _{{\rm{grid}}}^{k,j} \!=\! \left\{ {\begin{split}
&{\arcsin \left( {\frac{{(1 - \sqrt 2 /2)(2j - 1)}}{{2N}} - 1} \right) + \frac{{\pi (k - 1)}}{2},}\\
&{\arcsin \left( {\frac{{\sqrt 2 \left[ {2(j - N) - 1} \right]}}{{4N}} - \frac{{\sqrt 2 }}{2}} \right) + \frac{{\pi (k - 1)}}{2},}\\
&{\arcsin \left( {\frac{{(1 \!-\! \sqrt 2 /2)[2(j\!-\!3N) \!-\! 1]}}{{2N}} \!+\! \frac{{\sqrt 2 }}{2}} \right) \!+\! \frac{{\pi (k \!-\! 1)}}{2},}
\end{split}} \right.
\end{split}
\end{equation*}
\end{small}with piecewise $j = 1,...,N$, $j =N \!+\! 1,...,3N$, and $j =3N \!+\! 1,...,4N$, respectively. 
\begin{small}
\begin{equation*}
\begin{split}
&\theta _{{\rm{grid}}}^{k,l} = \left\{ {\begin{split}
&{\arccos \left( {\frac{{(1 - \sqrt 2 /2)(2l - 1)}}{{2N}} - 1} \right),}\\
&{\arccos \left( {\frac{{\sqrt 2 \left[ {2(l - N) - 1} \right]}}{{4N}} - \frac{{\sqrt 2 }}{2}} \right),}\\
&{\arccos \left( {\frac{{(1 - \sqrt 2 /2)[2(l-3N) - 1]}}{{2N}} + \frac{{\sqrt 2 }}{2}} \right),}
\end{split}} \right.\;\\
\end{split}
\end{equation*}
\end{small}with piecewise $l = 1,...,N$, $l =N \!+\! 1,...,3N,$, and $l =3N \!+\! 1,...,4N$, respectively. According to the proposed beams' distribution as well as their coverage shown in Fig. {\ref{stg}}, the set of grid directions/blocks covered by  $\bm{\omega} _i^{k,s}$ can be expressed as
\begin{subequations}
\begin{align}
&\Upsilon _i^{k,s} = \left\{ {(\phi _{{\rm{grid}}}^{k,j},\theta _{{\rm{grid}}}^{k,l})\;\left| {j \in \;{J_{s,i}},l \in {L_{s,i}}} \right.} \right\},\\
&{J_{s,i}} \!=\! \left\{ {N \!+\! \nu\cdot {{\bmod }_\mu }(i\!-\!1) \!+\! 1,...,N \!+\! \nu  ({{\bmod }_\mu }(i\!-\!1)\!+\!1)} \right\},\\
&{L_{s,i}} = \left\{ {N + \delta ({\rm{ceil}}\left( {\frac{i}{\mu }} \right) - 1) + 1,...,N + \delta \cdot {\rm{ceil}}\left( {\frac{i}{\mu }} \right)} \right\},\\
&\nu  = {2^{{\rm{ceil}}\left( {\frac{{S - s}}{2}} \right) + 1}},\;\;\delta  = {2^{{\rm{ceil}}\left( {\frac{{S - s - 1}}{2}} \right) + 1}},\;\;\mu  = {2^{{\rm{ceil}}\left( {\frac{{s - 1}}{2}} \right)}},\label{20d}
\end{align}
\end{subequations}  
where $\nu$ represents the number of grid blocks at the same elevation covered by $\bm{\omega} _i^{k,s}$, $\delta$ represents  the number of grid blocks at the same azimuth covered by $\bm{\omega} _i^{k,s}$, and $\mu$ represents the number of beams in stage $s$ at the same elevation.
Regarding the wide beams in stage $s$, prior works \cite{wide1,wide3,wide4} expect that $\bm{\omega} _i^{k,s}$ can only achieve beam gain within its coverage $\Upsilon _i^{k,s}$ and can not achieve gain in other range, i.e.,
\begin{equation}\label{dc}
{{\bf{a}}_k}{(\phi _{{\rm{grid}}}^{k,j},\theta _{{\rm{grid}}}^{k,l})^H}{\bm{\omega}} _i^{k,s} = \left\{ {\begin{split}
&{1,\;\;(\phi _{{\rm{grid}}}^{k,j},\theta _{{\rm{grid}}}^{k,l}) \in \Upsilon _i^{k,s}}\\
&{0,\;\;\;\;\;\;{\rm{otherwise}}\;\;\;\;\;}
\end{split}} \right.,
\end{equation}
holds true for all $j=1,2,...,4N$ and $l=1,2,...,4N$. However, it is worth mention that the feasible wide beam realized by the beamformer can not exactly fit (\ref{dc}), and only result in an approximate beam, which occurs notable trenches between adjacent ones. This is because the requirement of drastic jump/drop between $0$ and $1$ in (\ref{dc}) may squeeze the resulting beam patterns for minimizing the approximation error. These trenches bring dead zone and impair the overall performance of beam training.  To eliminate them, we propose to modify the criterion given in (\ref{dc}) by adding a buffer zone, which is effective and will be validated in Section \ref{BPC}. The buffer zone ${\rm B}_i^{k,s}$ is the periphery of $\Upsilon _i^{k,s}$ with width of $w$. To write it in mathematical form, we first set an enlarged zone of $\Upsilon _i^{k,s}$, denoted by $\hat\Upsilon _i^{k,s}$ as
\begin{equation}
\begin{split}
&\hat\Upsilon _i^{k,s} = \left\{ {(\phi _{{\rm{grid}}}^{k,j},\theta _{{\rm{grid}}}^{k,l})\;\left| {j \in \;{\hat J_{s,i}},l \in {\hat L_{s,i}}} \right.} \right\},\\
&{\hat J_{s,i}} = \{N + \nu\cdot {{\bmod }_\mu }(i-1)+1-w,...,\\
&\qquad\qquad\qquad\qquad\quad N + \nu  ({{\bmod }_\mu }(i-1)+1)+w \},\\
&{\hat L_{s,i}} = \left\{ N + \delta ({\rm{ceil}}\left( {\frac{i}{\mu }} \right) - 1)+1-w ,...,\right.\\
&\left.\qquad\qquad\qquad\qquad \quad N + \delta \cdot {\rm{ceil}}\left( {\frac{i}{\mu }} \right)+w \right\},\;(\ref{20d}).\\
\end{split}
\end{equation}
Then, we have ${\rm B}_i^{k,s}=\hat\Upsilon _i^{k,s}-\Upsilon _i^{k,s}$. Thus, in our proposed criterion, we expect that 
\begin{equation}\label{dc2}
{{\bf{a}}_k}{(\phi _{{\rm{grid}}}^{k,j},\theta _{{\rm{grid}}}^{k,l})^H}{\bm{\omega}} _i^{k,s} = \left\{ {\begin{split}
&{1,\;\;(\phi _{{\rm{grid}}}^{k,j},\theta _{{\rm{grid}}}^{k,l}) \in \Upsilon _i^{k,s}}\\
&{\chi ,\;\;(\phi _{{\rm{grid}}}^{k,j},\theta _{{\rm{grid}}}^{k,l}) \in {\rm B}_i^{k,s}}\\
&{0,\;\;\;\;\;\;{\rm{otherwise}}\;\;\;\;\;}
\end{split}} \right.,
\end{equation}
where $\chi \in (0,1)$ is the expected beam gain in the buffer zone. Define a matrix as
\begin{equation}
\begin{split}
&{\bf{A}} = \left[ {{{\bf{a}}_k}(\phi _{{\rm{grid}}}^{k,1},\theta _{{\rm{grid}}}^{k,1}),...,} \right.\\
&\qquad\qquad\left. {{{\bf{a}}_k}(\phi _{{\rm{grid}}}^{k,4N},\theta _{{\rm{grid}}}^{k,1}),...,{{\bf{a}}_k}(\phi _{{\rm{grid}}}^{k,4N},\theta _{{\rm{grid}}}^{k,4N})} \right]
\end{split}
\end{equation}
Then, we can rewrite (\ref{dc2}) in a more compact form as 
\begin{equation}
{{\bf{A}}^H}[{\bm{\omega}} _1^{k,s},{\bm{\omega}} _2^{k,s},...,{\bm{\omega}} _{{2^s}}^{k,s}] = {\bm{\Xi} _s}.
\end{equation}
where ${\bm{\Xi} _s}$ is a $16N^2 \times 2^s$ matrix, which has an element of 1 in the coverage  zone,  an element of $\chi$ in the dead zone, and an element of $0$ in other zone. As a result, the $i$th codeword in the stage  $s=0,1,...,S-1$ of $\mathcal{C}_k$ can be obtained as 
\begin{equation}\label{wide}
{\bm{\omega}} _i^{k,s} = {({\bf{A}}{{\bf{A}}^H})^{ - 1}}{\bf{A}}{\bf{\Xi}} _s(:,i).
\end{equation}

Now, we have obtained all the codewords in $\mathcal{C}_k$. The narrow beams in stage $S$ are given by (\ref{narrow}) and the wide beams in stage $s=0,1,...,S-1$ are given by (\ref{wide}).
\subsection{3D Training Procedure}
\begin{figure}[t]
\centering
\includegraphics[width=3.5in]{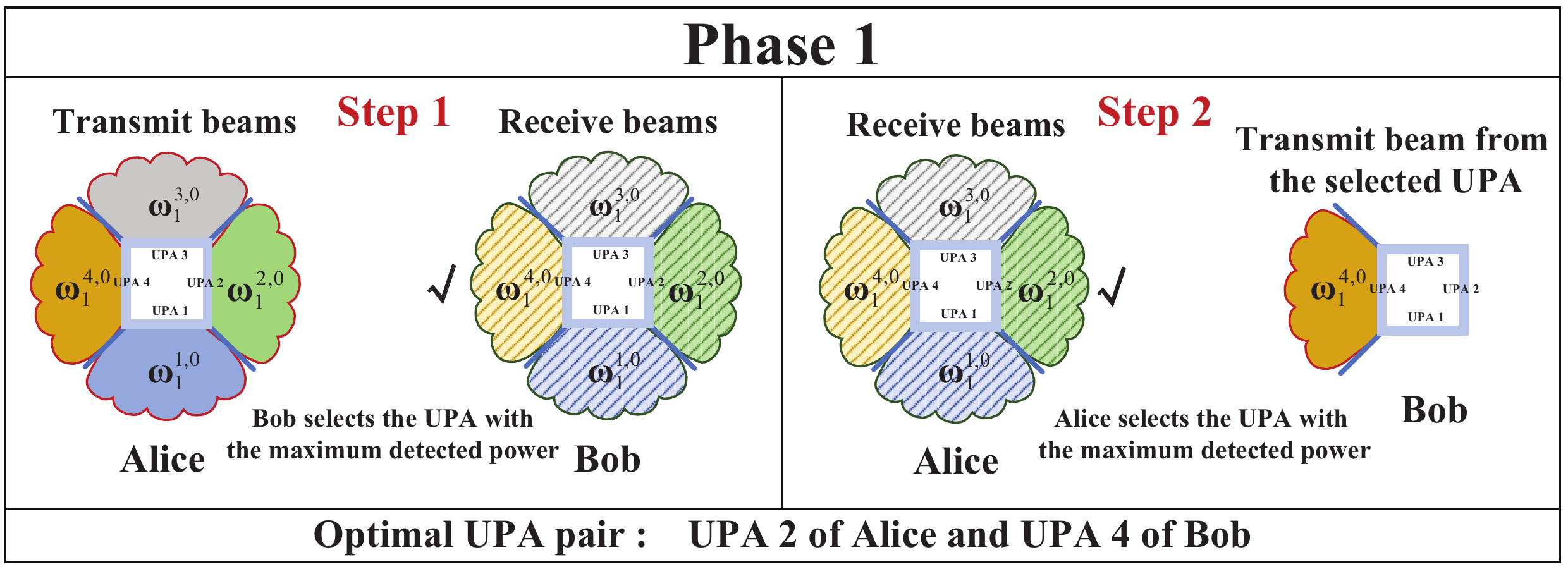}
\caption{Phase 1 of the beam training procedure shown on the $xy$-plane.}\label{ph1}
\vspace{-12pt}
\end{figure}
In this subsection, we propose a low-complexity 3D training procedure for our considered system. We name the two terminals as Alice and Bob respectively. To find the optimal narrow-beam pair between Alice and Bob, two phases are developed to achieve different groups of measurements. Due to the page limit, we omit the details in the description, and the overall procedure is sketched in Fig. {\ref{ph1}} and Fig. {\ref{ph2}}.
 \subsubsection{Phase 1 (see Fig. {\ref{ph1}})}
\begin{figure}[t]
\centering
\includegraphics[width=3.5in]{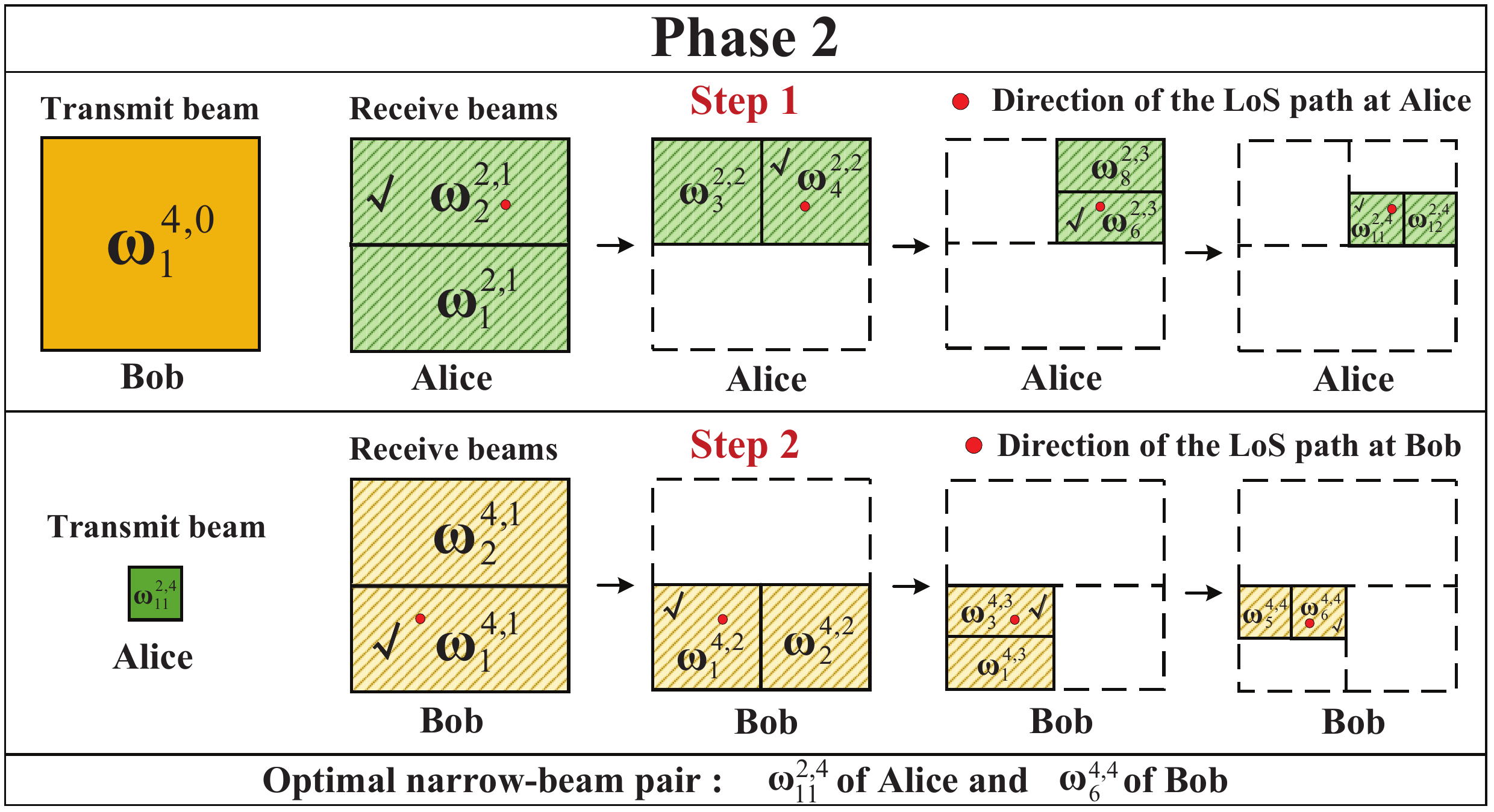}
\caption{Phase 2 of the beam training procedure shown on $\Theta (\theta )$ and $\Phi (\phi )$, where $A^*=2$ and $B^*=4$. }\label{ph2}
\vspace{-12pt}
\end{figure}

In Phase 1, we aim to find the optimal UPA pair whose beam range covers the LoS path.  In step 1, Alice simultaneously uses all UPAs to transmit wide beams. In the meanwhile, Bob simultaneously uses all UPAs to receive wide beams. After the measurement, Bob selects the one with the maximum detected power. In step 2, Bob only transmits the wide beam by the selected UPA. In the meanwhile, Alice simultaneously uses all UPAs to receive wide beams. After the measurement, Alice finds the UPA (labeled as $A^*$) with the maximum detected power in the same way. After the two steps, the optimal UPA pair is obtained as the $A^*$th UPA of Alice and the $B^*$th UPA of Bob.
\subsubsection{Phase 2 (see Fig. {\ref{ph2}})}
In Phase 2, we aim to find the optimal narrow-beam pair between $A^*$th UPA of Alice and the $B^*$th UPA of Bob. In step 1, Bob transmits a wide beam using the $B^*$th UPA. In the meanwhile, Alice uses the $A^*$th UPA to receive wide beams via testing some codewords from stage $1$ to stage $S$. In each stage, Alice tests two beams and selects the one with larger detected power, and in the next stage, Alice tests two beams that are within the range of the selected beam. By recursively repeating this way, Alice can reach a narrow beam (labeled as $a^*$) in the stage $S$. In step 2, Alice transmits a narrow beam using the $A^*$th UPA. In the meanwhile, Bob uses the $B^*$th UPA to hierarchically test codewords by the same way, and reach a narrow beam (labeled as $b^*$) in the stage $S$. After the two steps, the optimal narrow-beam pair is obtained as ${\bm{\omega}} _{a^*}^{A^*,S}$ of Alice and ${\bm{\omega}} _{b^*}^{B^*,S}$ of Bob.

\section{Numerical Results}\label{nu}
In this section, numerical results are provided to demonstrate the beam patterns of our proposed hierarchical codebook and the performance of our proposed beam training.
\subsection{Beam Patterns of the narrow beams}
\begin{figure}[t]
\centering
\includegraphics[width=2.8in]{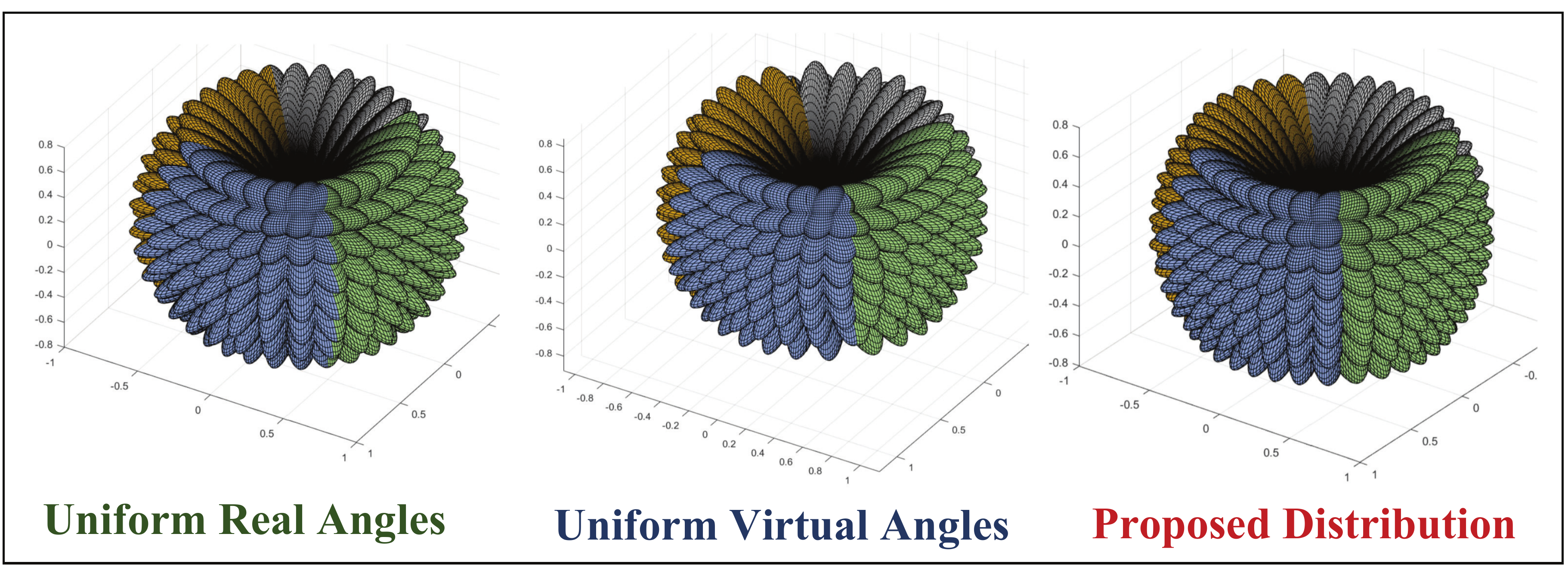}
\caption{Comparison of the proposed narrow beams with the benchmarks, where $N_y=N_z=8$ and $N=8$.}\label{nacom}
\vspace{-12pt}
\end{figure}
We first consider the beam patterns of our proposed narrow beams, which are in the bottom stage of the hierarchical book. For comparison, we present two benchmarks as follows.
\begin{itemize}
\item {\emph{Uniform Real Angles:}} For the $k$th UPA, we extend the method in \cite{ywang} by setting $N$ azimuth angles uniformly distributed in $[ - \frac{\pi }{4} + (k - 1)\frac{\pi }{2},\frac{\pi }{4} + (k - 1)\frac{\pi }{2}]$ and $N$ elevation angles uniformly distributed in $[\frac{\pi }{4},\frac{3\pi }{4}]$.
\item {\emph{Uniform Virtual Angles:}} For the $1$th UPA, we consider the simplified array response vector with virtual angles (also named spatial angles), i.e., 
\begin{equation}
\begin{split}
&{{\bf{a}}_1}(\widetilde \phi ,\widetilde \theta ) = \frac{1}{{\sqrt {{N_a}} }}[1,...,{e^{j\pi ({n_y}\widetilde \phi  + {n_z}\widetilde \theta )}},\\
&\qquad\qquad\qquad\qquad...,{e^{j\pi [({N_y} - 1)\widetilde \phi  + ({N_y} - 1)\widetilde \theta ]}}]^T,
\end{split}
\end{equation}
where $\widetilde \phi$ and $\widetilde \theta$ are the virtual azimuth and elevation angles within $[-1,1]$. We set $N$ virtual azimuth angles and $N$ virtual elevation angles uniformly distributed $[ - \frac{{\sqrt 2 }}{2},\frac{{\sqrt 2 }}{2}]$. For the $k$th UPA ($k=2,3,4$), we rotate the beam patterns of the $1$th UPA $(k - 1)\frac{\pi }{2}$ in azimuth.
\end{itemize}
Fig. \ref{nacom} plots the narrow-beam patterns of our proposed distribution and the benchmarks, where each UPA has $8\times 8$ antenna elements and uses $8\times 8$ narrow beams to cover its range. It is observed that the trenches of the narrow beams with uniform real angles are deeper than those with our proposed distribution. Note that the azimuth coverage of the narrow beams of benchmarks expand at high elevation angle, which cause a certain overlap between adjacent UPAs. This indicates that our proposed narrow beams have better worst-case performance, and yield more uniform beam patterns, especially for the beams at the edge of the coverage of each UPA. 
\subsection{Beam Patterns of the Wide beams}\label{BPC}
Next, we consider the beam patterns of our proposed wide beams $\{\bm{\omega} _i^{k,s}\}_{s=0}^{S-1}$, which are in the stage $0$ to $S-1$ of the hierarchical book. For comparison, we extend the inverse approach adopted in \cite{wide1} and \cite{wide4} to the 3D scenario as a benchmark, the wide beam $\bm{\omega} _i^{k,s}$ are designed for covering the range $\Upsilon _i^{k,s}$.
\begin{figure}[t]
\centering
\includegraphics[width=3.5in]{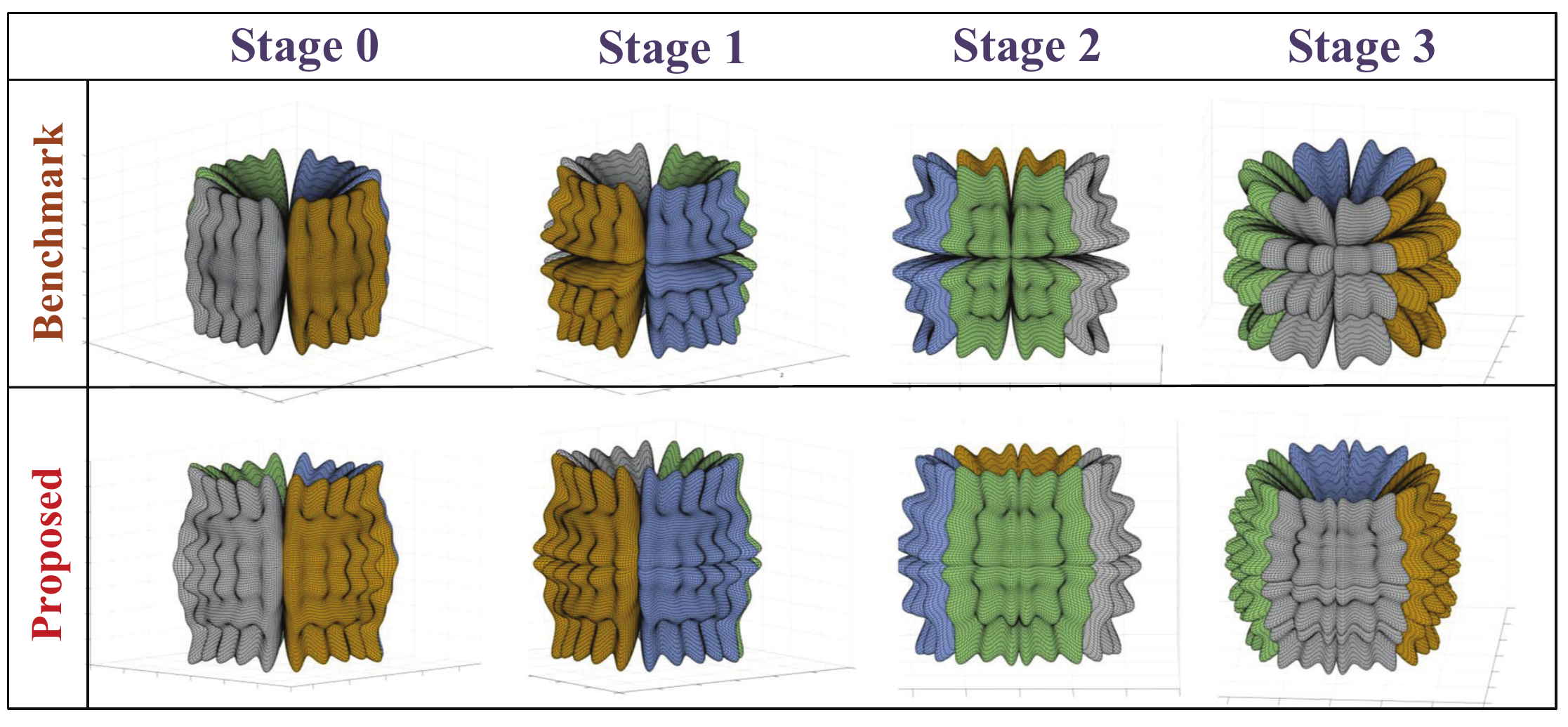}
\caption{Comparison of the proposed wide beams with the benchmarks, where $N_y=N_z=16$ and $N=16$.}\label{wicom}
\vspace{-10pt}
\end{figure}
Fig. \ref{wicom} plots the wide-beam patterns in stage $0$ to $3$ of our proposed codebook and the benchmark, where each UPA has $16\times 16$ antenna elements and the codebook has $16\times 16$ narrow beams in the bottom stage. It is observed that in stage $0$, the wide beams realized by both approaches all have notable trenches between adjacent UPAs. In contrast, the trenches of our proposed wide beams are relatively smaller. In stage $1$ to $3$, the wide beams realized by the benchmark have remarkable trenches even within the coverage range of each UPA. By comparison, there are no trenches within it in the patterns of our proposed wide beams. The beam patterns imply that using our proposed wide beams will have less dead zone during the beam training, and thus are expected to achieve better performance.
\subsection{Performance of Beam Training}
\begin{figure}[t]
\centering
\includegraphics[width=2.6in]{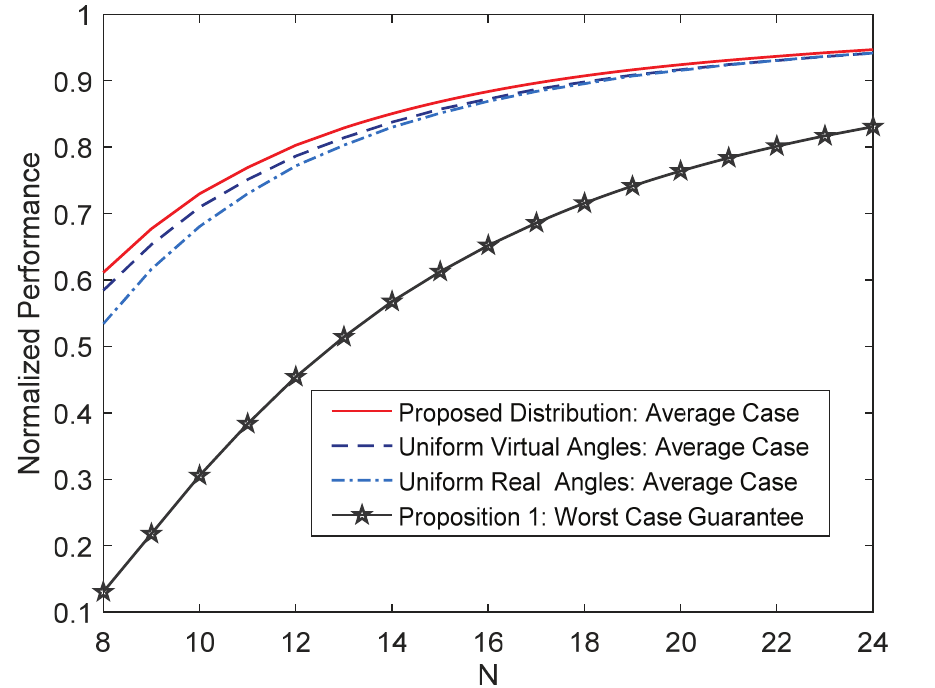}
\caption{Normalized performance versus number of narrow beams, where $N_y=N_z=16$.}\label{ng}
\vspace{-10pt}
\end{figure}

To evaluate the performance of the hierarchical beam training, we consider looking into the average/worst-case performance and the successful alignment rate. With a successful alignment, the final performance is determined by the narrow beams in the bottom stage. Fig. {\ref{ng}} shows the normalized performance, i.e., normalized beam gain, of using different narrow beams, in which the worst-case performance is presented as a baseline. As can be seen, the performance of all schemes increases with the number of narrow beams, and our proposed narrow beams outperform the benchmarks.

\begin{figure}[t]
\centering
\includegraphics[width=2.6in]{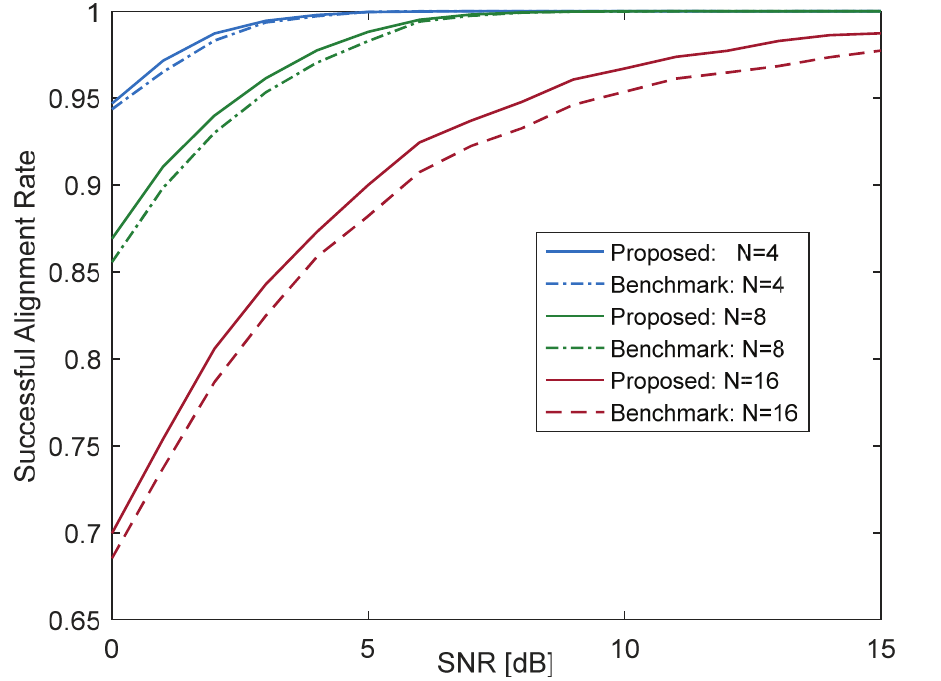}
\caption{Successful alignment rate versus SNR, where $N_y=N_z=8$.}\label{sar}
\vspace{-12pt}
\end{figure}

Fig. \ref{sar} shows three sets of results (denoted by different colors,
respectively) of the successful alignment rate by using our proposed hierarchical codebook as well as the benchmark shown in Fig. \ref{wicom}. It can be observed that both schemes can achieve 100\% successful alignment rate with sufficiently high SNR, and our proposed codebooks can outperform the benchmark codebook in different setups. Thus, our proposed scheme can find a high-quality beamforming solution, i.e., the optimal narrow-beam pair, even in the severe-path-loss THz channel, and the performance is comparable compared to optimal beamforming.

\section{Conclusions}
We considered a quadruple-UPA architecture for THz communication. Considering the severe path loss of pilot signals propagation, we propose a fast 3D beam training strategy to jointly realize AoD/AoA estimation and beamforming. To this end, we first develop a novel hierarchical codebook to achieve optimal spatial coverage and then propose a 3D beam training procedure to find the optimal narrow-beam pair quickly. Numerical results visually plot the beam patterns of our proposed codebook and validate the performance of the proposed beam training.

\end{document}